\documentclass[runningheads]{llncs}

\usepackage{graphicx}
\usepackage{multirow}
\usepackage{subcaption}
\usepackage{xcolor,pifont}
\usepackage{array}
\usepackage{footnote}
\begin{document}
\title{A Large-scale Multi Domain Leukemia Dataset for the White Blood Cells Detection with Morphological Attributes for Explainability}
%
%
\author{Abdul Rehman\inst{1} \and
Talha Meraj\inst{1} \and
Aiman Mahmood Minhas \inst{2} \and
Ayisha Imran \inst{2} \and
Mohsen Ali \inst{1} \and
Waqas Sultani\inst{1}
}
%
\authorrunning{A. Rehman et al.}

\institute{Intelligent Machines Lab, Information Technology University, Lahore, Pakistan
\email{\{phdcs23002,talha.meraj,mohsen.ali,waqas.sultani\}@itu.edu.pk}
\and
Department of Hematology, Chughtai Labs, Lahore, Pakistan
}

\maketitle          
\begin{abstract}
Earlier diagnosis of Leukemia can save thousands of lives annually.  
The prognosis of leukemia is challenging without the morphological information of White Blood Cells (WBC) and relies on the accessibility of expensive microscopes and the availability of hematologists to analyze Peripheral Blood Samples (PBS).
 Deep Learning based methods can be employed to assist hematologists. However, these algorithms require a large amount of labeled data, which is not readily available.  
To overcome this limitation, we have acquired a realistic, generalized, and {large} dataset.  
To collect this comprehensive dataset for real-world applications, two microscopes from two different cost spectrums (high-cost: HCM and low-cost: LCM) are used for dataset capturing at three magnifications (100x, 40x,10x) through different sensors (high-end camera for HCM, middle-level camera for LCM  and mobile-phone's camera for both). The high-sensor camera is 47 times more expensive than the middle-level camera and HCM is 17 times more expensive than LCM. In this collection, using HCM at high resolution (100x), experienced hematologists annotated 10.3k WBC types (14) 
and artifacts, having 55k morphological labels (Cell Size, Nuclear Chromatin, Nuclear Shape, etc) from 2.4k images of several PBS leukemia patients. 
Later on, these annotations are transferred to other 2 magnifications of HCM, and 3 magnifications of LCM, and on each camera captured images. 
Along with the LeukemiaAttri dataset, we provide baselines over multiple object detectors and Unsupervised Domain Adaptation (UDA)  strategies, along with morphological information-based attribute prediction. 

The dataset will be publicly available after publication to facilitate the research in this direction.

\end{abstract}
\section{Introduction}
\label{sec:intro}
According to GLOBOCAN  2020, Leukemia is a leading cause of cancer-related deaths in individuals under 39 years, especially children. 
It constitutes 2.5\% of total cancer incidences with an annual estimate of 474,519 cases, leukemia is a rare yet highly malignant disease  \cite{chhikara2023global}.
Leukemia, a form of hematologic malignancy, presents a frightening challenge in modern medicine due to its diverse subtypes, complex etiologies, and varying disease progressions \cite{gbenjo2023leukemia}. 
Initiated through genetic mutations in the bone marrow cells, disrupting the normal development and count of various blood cells, leading to uncontrolled growth of abnormal malignant WBC \cite{walkovich2022disorders}.
 Conventional methods for diagnosing leukemia often involve specialized laboratory tests, demanding extensive sample preparation and expensive medical equipment \cite{shah2021automated}. Particularly in remote regions of developing countries, the management of leukemia faces challenges arising from the scarcity of costly laboratory equipment and, notably, a shortage of trained technicians and specialized doctors  \cite{rugwizangoga2022experience}.
\begin{figure*}[!t]
    \includegraphics[width=\linewidth]{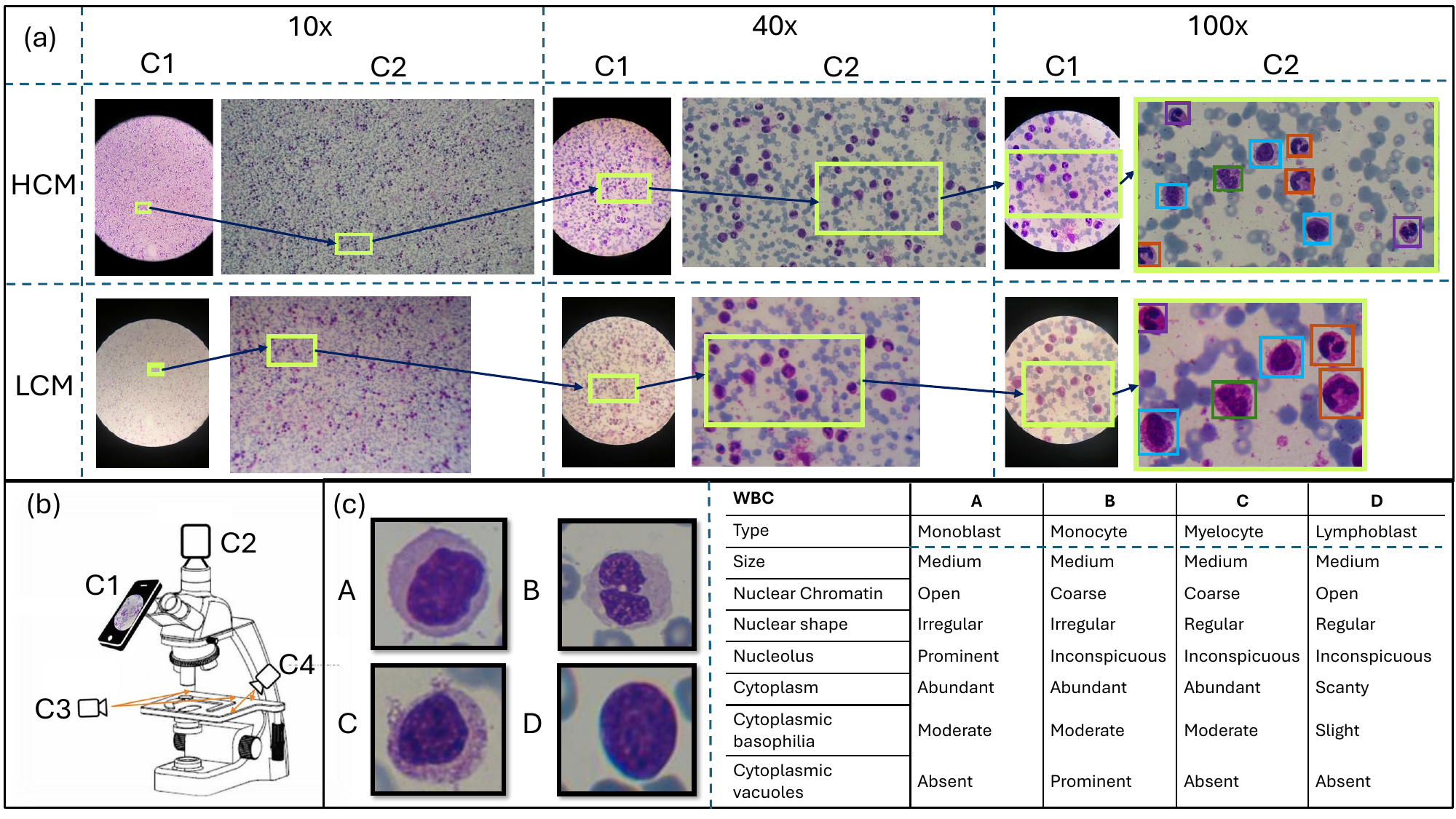}
    \caption{a) Illustrate the image-capturing procedure using general mobile cameras (C1) and high-end cameras for HCM, and middle-level cameras for LCM (C2) at multiple resolutions. Images are captured using both high and low-cost microscopes. b) shows our microscope set up with cameras (C1, C2) to capture slide images  and cameras (C3, C4) to capture the stage-scales. c) Shows the different types of WBCs with morphological attributes.
    }
    \label{fig:1}
\end{figure*}
Precise and fast diagnosis is necessary for timely initiation of appropriate treatment, extremely influencing the survival of patients \cite{azad2015short}.
 However, less availability of expensive medical equipment makes it necessary to enable low-cost equipment for diagnostic purposes \cite{mallya2022deep}.  In clinical practices, the microscopic examination of Peripheral Blood Film (PBF) is a very first step for the leukemia diagnosis. The choice of microscope and its resolution, along with the training of the medical practitioners, affects the accuracy of such diagnosis, for example, identifying the WBC's type is feasible at a 40x resolution but becomes challenging at 10x.   
However, for detailed analysis of cell morphology to ensure an accurate prognosis for leukemia, a high-quality microscope and higher resolution of about 100x is preferred. 
Thus,  the PBF analysis for the prognosis of leukemia is a knowledge-intensive and expensive process, necessitating the use of costly microscopes and trained experts. 
Similarly, cost-effective diagnostic modalities often lacking in low-resource areas \cite{manescu2022automated}.
The limitation of these factors collectively restricts accessibility to early and accurate leukemia prognoses, particularly in remote and resource-constrained areas. 
To address the aforementioned factors, subjectivity, and the shortage of hematologists, Artificial Intelligence (AI), especially deep learning-based methods has been recently proposed, along with datasets.
However, it is crucial to acknowledge that previously published datasets ~\cite{scotti2005automatic,labati2011all,rezatofighi2011automatic,gupta2019all,matek2019single,aria2021acute,kouzehkanan2022large,bodzas2023high} lack in various aspects, some are limited by several samples, others do not attend to the problem of localization of the WBC, many do not have information about the morphological attributes and most are only over one sensor or microscope,  etc. All of these limit the development of a solution that could be applied in real-world scenarios.
To tackle the above-mentioned challenges and to assist the hematologists with an explainable second opinion on the prognosis of leukemia, a morphological information enriched, multi-domain large-scale images dataset, namely LeukemiaAttri is collected.
The collection of the LeukemiaAttri dataset, consisting of 28.9K images captured using low-cost and high-cost microscopes at three different resolutions (10x,40x,100x) and different cameras.  
In addition to the location annotation of each WBC, we provide detailed morphological attributes for each WBC. The attributes include WBC size, nuclear chromatin, nuclear shape, nucleolus, cytoplasm, cytoplasmic basophilia, and cytoplasmic vacuoles. 
These attributes were selected after a conversation with multiple hematologists. The procedure of dataset collection is illustrated in Fig. \ref{fig:1}.

In existing CAD systems, many WBC detectors have been employed, leveraging methods Faster-RCNN, YOLOv5, and other object detectors \cite{sun2021sparse,tian2019fcos,zhang2022dino,ultralytics2021yolov5}. Although, these object detectors offer satisfactory solutions for WBC detection but lack explainability, vital for the leukemia's prognosis. To overcome this limitation, we have provided a multi-head object detector approach, namely AttriDet, that not only detects the WBC types but also predicts their morphological attributes employing low-level and highly deep features. We are hopeful that our work will assist the hematologists in providing a more confident prognosis. In addition, we have provided several competitive baseline results of state-of-the-art object detection and UDA methods. The dataset will be made publicly available. 
To summarise our contributions; {(1)} A large-scale multi-domain WBC detection benchmark along with morphological attributes of WBCs for prognosis of leukemia is introduced\footnote{morphological attributes, recommended by hematologists for prognosis of leukemia},
{(2)} To facilitate future research, we have constructed extensive WBC's detection and UDA baselines, 
{(3)} A multi-headed WBC detection and morphological attribute prediction architecture are introduced.

\section{Dataset}
\label{sec:Methods}
 
{Popular datasets cover four types of leukemia including Acute Lymphocytic Leukemia (ALL), Acute Myeloid Leukemia
(AML), Chronic Lymphocytic Leukemia (CLL), and Chronic Myeloid Leukemia (CML).   These types mainly fall into two lineages namely myeloid (AML and CML) and lymphoid (ALL and CLL). Specific characteristics include increased CML myeloblasts, atypical CLL lymphocytes, elevated AML myeloblasts, and ALL lymphoblasts.
\begin{table}[h!]
  \centering
    \caption{Comparison of the proposed dataset with existing leukemia datasets.}  
    \label{tab:1}
\begin{tabular}{|l|l|c|c|c|c|l|l|c|}
    \hline
    &&Across&Multi. Cells&BBX&Multi.&No. of & WBC& Morphology \\
    Dataset&Type&Micro.&in image&&Res.&WBC's&Classes&\\
    \hline
     IDB \cite{scotti2005automatic} & ALL&\ding{56}
     & \textcolor{green}{\ding{52}} 
&\textcolor{green}{\ding{52}}&\ding{56}&510 (LB)&2& 
        \ding{56}\\
        IDB2 \cite{labati2011all} & ALL & \ding{56}& 
        \ding{56}
         & \ding{56} &\ding{56}&260&2& 
        \ding{56}\\
        LISC \cite{rezatofighi2011automatic}& Multi. & \ding{56}
        &\ding{56} 
        & \ding{56}&\ding{56}&  250&6& 
        \ding{56}\\
        Munich\cite{matek2019single} & AML & \ding{56}
        & \ding{56}
        & \ding{56}&\ding{56}&18,365&15& 
        \ding{56} \\
       Raabin \cite{kouzehkanan2022large}& Normal & \ding{56} & \textcolor{green}{\ding{52}}& \textcolor{green}{\ding{52}}&\ding{56}& 
     17,965& 5& 
        \ding{56}\\
        HRLS \cite{bodzas2023high}   & Multi. & \ding{56} &\textcolor{green}{\ding{52}}
       & \ding{56}&\ding{56} &16,027&9& 
        \ding{56}\\

         WBCAtt \cite{tsutsui2024wbcatt}   & Normal & \ding{56} &\ding{56}
       & \ding{56}&\ding{56} &10,298&5& 
        \textcolor{green}{\ding{52}}\\
        
     \textbf{Ours} \footnotemark &Multi.&\textcolor{green}{\ding{52}}
     &\textcolor{green}{\ding{52}}&\textcolor{green}{\ding{52}}&\textcolor{green}{\ding{52}}
  &\textbf{88,294}&14& 
        \textcolor{green}{\ding{52}}\\
    \hline
  \end{tabular}
\end{table}
 \footnotetext{The details LeukemiaAttri is explained in supplementary material section}
As shown in  Table \ref{tab:1}, most existing WBC datasets were collected using a single microscope, without localization and morphology whereas most contain individuals or a single type of leukemia patient. Furthermore, they lack localization and explainability regarding the WBC morphology of leukemia patients which is present in our dataset.  
\subsection{LeukemiaAttri Dataset} 
To gain a comprehensive understanding of the Leukemia prognosis and its impact, we discussed with several healthcare professionals from different working environments and finalized the WBC types and their morphological attributes.
\begin{figure}[!htbp]
    \includegraphics[width=\linewidth]{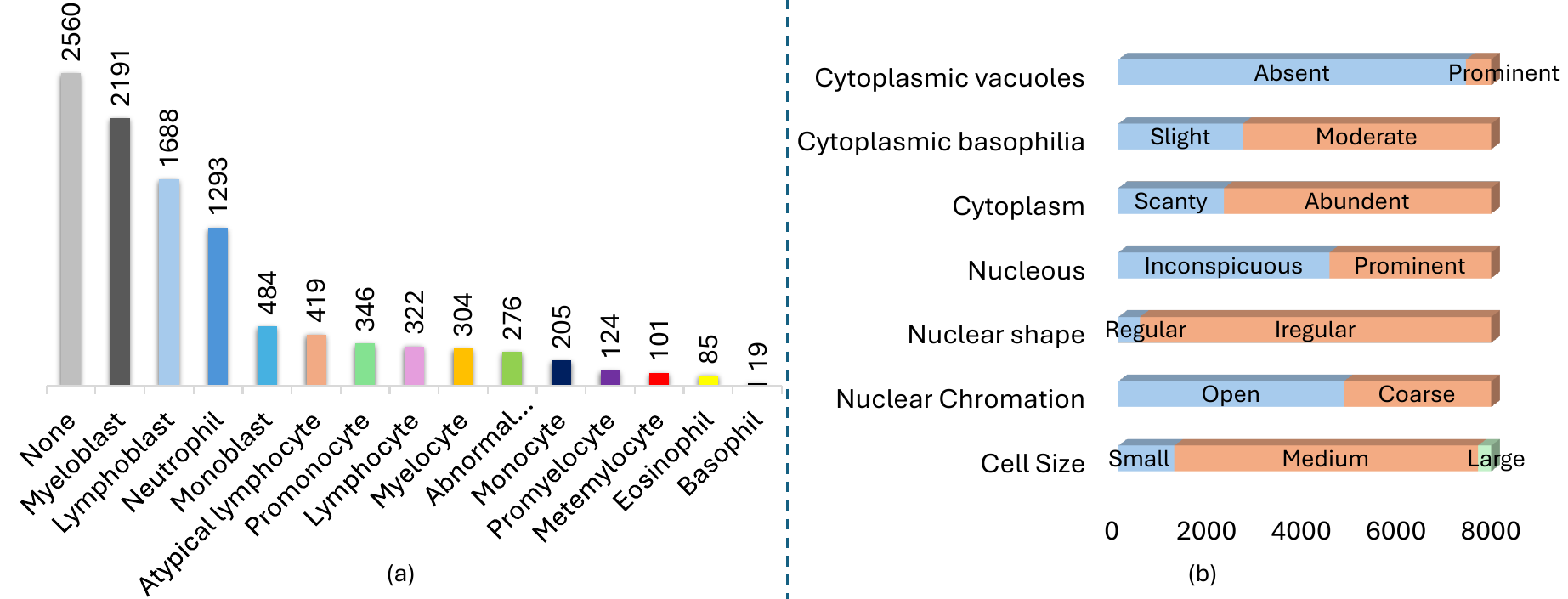}
    \caption{(a): WBC type and (b) Morphological attributes distribution}
    \label{fig:2}
\end{figure}
In this dataset collection, the PBFs are utilized from the diagnostic lab, ensuring patient consent, and incorporating the hematologist's marks in the region from the monolayer area. To capture images, we utilize two distinct microscopes – the high-cost (Olympus CX23) and the low-cost (XSZ-107BN) – in conjunction with two separate cameras, namely the HD1500T (HCM), ZZCAT 5MP (LCM) and the Honor 9x Lite mobile camera (HCM, LCM). It is quite challenging to locate the same patch on the PBF when employing different microscopes and resolutions \cite{sultani2022towards}. To address this inherent challenge, we initiated the capturing process by setting the field of view (FoV) at 10x, and 40x with an approximate 20\% overlap, maintaining a fixed x-axis stage scale. 
At 100x magnification, we captured the FOV containing WBCs without any overlap, ensuring the distinct representation of individual WBCs. This process was repeated both for HCM and LCM.

\noindent\textbf{Morphological attributes:} The set of rules for WBC morphology varies depending on the hematologists. To enhance prognostic assistance, hematologists identified the 14 types of WBC and considered seven key morphological attributes for a well-informed prognosis. To annotate the WBC type and morphology attributes, hematologists reviewed the subsets of the captured images and then selected the most structural and best quality images from the given subset (HD1500T paired camera at 100x using HCM) of the LeukemiaAttri dataset. For quality control, two hematologists annotate each cell with the consultation. The detail of some types of WBC with the morphological information is shown in  {the Fig. \ref{fig:1} section (c), where A) monoblasts, B) monocytes, and C) myelocytes cells exhibit mostly similar morphological characteristics as they originate from the myeloid lineage. Nevertheless, differences arise, particularly in the presence of cytoplasmic vacuolation. However, D) lymphoblasts belong to a lymphoid lineage that shows morphological dissimilarities in both lineages.} After obtaining detailed WBC and their attributes annotations from hematologist for HCM at 100x, we transferred the annotations to different resolutions and across microscope automatically using homography \cite{lowe2004distinctive}, \cite{fischler1981random}. Transferred annotations were verified manually and re-annotatation was done for the missing localization. The detailed count of the source subset of WBC types and their corresponding attributes are shown in Fig. \ref{fig:2} (please see supplementary material for details).
\begin{figure}[!htbp]
    \includegraphics[width=\linewidth]{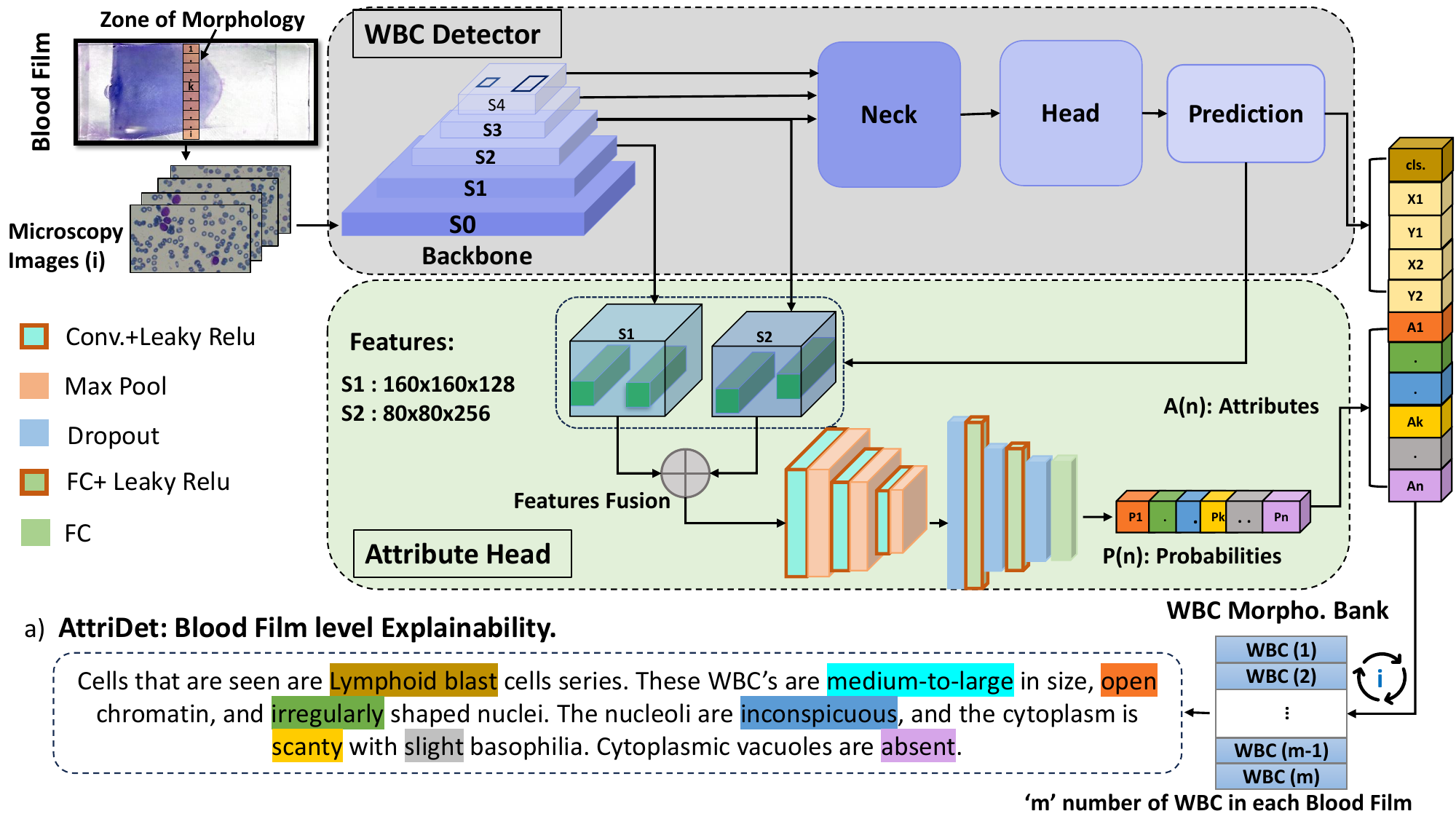}
    \caption{AttriDet: WBC and their corresponding attribute predictions framework}
    \label{fig:3}
\end{figure}
\section{AttriDet: A WBC detection method with attributes}
Although there have been several approaches presented recently to detect WBC \cite{zhang2022rcmnet,manescu2023detection}. However, a clear lack of explainable WBC detection is observed \cite{tsutsui2024wbcatt}. 
\noindent\textbf{AttriDet:} To achieve explainable WBC detection, we propose to use a multi-headed WBC detector.  We firstly apply recent object detectors from different domains including one-stage (FCOS \cite{tian2019fcos}, YOLOv5\cite{ultralytics2021yolov5}), two-stage (Spare-Faster-RCNN \cite{sun2021sparse}) detectors, and transformer (DINO\cite{zhang2022dino}). We chose these methods because of their good detection results, efficiency, and low memory consumption. Due to the overall better result of YOLOv5 on our datasets (see Table \ref{tab:2}), we have extended the YOLOv5 for attribute prediction. In YOLOv5 architecture, we added a lightweight attribute head for the prediction of WBC's morphology as shown in Fig. \ref{fig:3}. 
For attribute prediction, we want to capture low-level visual details, therefore, we fuse features from two initial layers which caries structural and semantically enriched information.
To train the attribute head with YOLOv5 heads, asymmetric loss \cite{ridnik2021asymmetric} is employed. The YOLOv5 method-based object detections and attribute head prediction collectively predict the cell and its morphology which gives the explainable reasoning. 
These predictions have been registered in a WBC morphology bank and blood film level descriptions have been generated based on the most frequently appearing WBC type (recommended by hematologists). An example of a generated text-based description is shown in Fig. \ref{fig:3} (a).  
We discussed AttriDet produced explainability with the hematologists who appreciated this explanation and see our method as a potential candidate for a second opinion of leukemia prognosis. Furthermore, the presented AttriDet not only predicts associated attributes to a cell but also increases the YOLOv5 predictability (improved mAP from 26.3 to 28.2 as shown in Table \ref{tab:2}, and \ref{tab:4} on subset H\_100x\_C2)

\section{Experiments}
\subsection{Object Detection}
The LeukemiaAttri dataset contains 14 types of WBC cells with 1 class of artifacts in the None category.  For object detection experiments, we have used four subsets from the LeukemiaAttri dataset, namely H\_100x\_C1 (Mobile), H\_100x\_C2 (HD1500T), L\_100x\_C1 (Mobile), and L\_100x\_C2 (MZZCAT 5MP). The experimental results are shown in  Table \ref{tab:2}. 
\setlength{\tabcolsep}{8pt} 
\begin{table}[h]
\caption{Object Detection baselines results on LeukemiaAttri dataset}
\label{tab:2}
\centering
\begin{tabular}{|c|c|c|c|}
\hline
\multirow{1}{*}{Method}\hspace{1.5em} &\multirow{1}{*} {Subset} \hspace{1.5em}& {mAP 50-95}\hspace{0.5em}&  \multirow{1}{*}{mAP50}\hspace{0.5em} \\
\hline
\multirow{4}{*}{Sparse R-CNN \cite{sun2021sparse}\hspace{1.5em}} & H\_100x\_C1  &15.8\hspace{1.5em}&29.6\hspace{1.5em}\\
 & H\_100x\_C2 &21.3 & 36.7 \\
 & L\_100x\_C1   &17.2   & 32.6 \\
 & L\_100x\_C2   &14.5 & 25.9 \\
\hline
\multirow{4}{*}{FCOS \cite{tian2019fcos} } & H\_100x\_C1   &16.4&31.8 \\
 & H\_100x\_C2 & 22.5 & 40.6 \\
 & L\_100x\_C1   &17.5&33.9 \\
& L\_100x\_C2  &17.7&34.3 \\
\hline
\multirow{4}{*}{DINO\cite{zhang2022dino}}& H\_100x\_C1  &17.0&33.8 \\
  & H\_100x\_C2  &25.4& 43.7\\
 & L\_100x\_C1   &17.5&34.3 \\
 & L\_100x\_C2   &21.5&38.2\\
\hline
\multirow{4}{*}{YOLOv5x\cite{ultralytics2021yolov5}}& H\_100x\_C1  &\textbf{20.9}&\textbf{38.8} \\
  & H\_100x\_C2  &
\textbf{26.3} & \textbf{44.2}\\
 & L\_100x\_C1  &\textbf{20.7}&\textbf{39.5} \\
 & L\_100x\_C2  &\textbf{20.1}&\textbf{38.1} \\
\hline
\end{tabular}
\end{table}
According to the results, on the HCM100x\_C1 subset, the highest mAP50-95 was achieved by YOLOV5x 20.9 mAP50-95 and 33.8 mAP50. Similarly, on the HCM100x\_C2 subset, the highest results were achieved by YOLOv5x with 26.3 mAP50-95 and 44.2 mAP50. A similar pattern can be observed for other subsets as well. We believe this is due to the YOLOV5 robust feature learning for different sizes of cells presented in the LeukemiaAttri dataset object sizes. Since, collectively, on HCM and LCM subsets of microscope and mobile cameras, the YOLOv5x method showed higher results, we extend this for attribute prediction.


\subsection{Unsupervised Domain Adaptation based Object detection}

In the LeukemiaAttri dataset, 12 subsets are collected from different domains via HCM and LCM. These different domain subsets contain challenging domain shifts as shown in Fig. \ref{fig:1}: sec(a), making our dataset a domain adaption benchmark. Note that due to the poor image quality of LCM, the hematologists are usually reluctant to provide annotation as it is tedious and susceptible to errors. Therefore, UDA methods could be used to train classifiers on high-quality images of precise annotation (HCM) and provide results on LCM.
\setlength{\tabcolsep}{3pt} 
\begin{table}[h]
\caption{Object Detection based domain adaptations results on H\_100x\_C2 and L\_100x\_C2 subsets of LeukemiaAttri dataset}
\label{tab:3}
\centering
\begin{tabular}{|c|c|c|c|c|}
\hline
\multirow{1}{*}{Method} & {Train Subset } & {Test Subset} & {mAP50-95}&  \multirow{1}{*}{mAP-50} \\
\cline{2-4}
\hline
YOLOv5\cite{ultralytics2021yolov5} (source only) & H\_100x\_C2 & L\_100x\_C2  &11.0&25.5\\ \hline
DACA \cite{mekhalfi2023daca} 
 & H\_100x\_C2  & L\_100x\_C2 &12.6 &30.2\\ \hline
ConfMix \cite{sun2021sparse}  & H\_100x\_C2 & L\_100x\_C2  &12.6&33.5\\ \hline
\end{tabular}
\end{table}
To provide the baselines of UDA, we have experimented with two recent methods; \cite{sun2021sparse} and DACA \cite{mekhalfi2023daca},  utilizing the higher resolution (100x) subsets collected via HCM and LCM. As can be seen in Table \ref{tab:3}, YOLOv5x (source only) was trained on the HCM subset, achieving a 25.5 mAP50 score on a comparable subset of the LCM. Nevertheless, employing UDA methods such as ConfMix and DACA led to higher mAP50 scores of 33.5 and 30.2, respectively. The low performance of state-of-the-art UDA methods demonstrates the complexity and large domain shift in our dataset.  




\subsection{Object detection with attribute prediction }

Table. \ref{tab:4} demonstrate results our proposed for AttriDet for different attributes prediction. In addition, we  show the improved WBC detection of AttriDet as compared to standard YOLOv5. We believe that better WBC results are due to robust feature learning employing attribute head. The results of CBM \cite{keser2023interpretable}  and AttriDet show that the proposed AttriDet is more efficient and confident in predicting the WBC types and their associated attributes. However, according to the results of CBM and AttriDet, the nucleus and cytoplasmic vacuoles have proven to be the most difficult attributes to detect.


\begin{table*}[ht]
\center
\caption{Testing set results of AttriDet and SOTA methods on H\_100x\_C2 subset:
WBC Type, Attributes (NC: Nuclear Chromatin,	NS: Nuclear shape, 	N: Nucleus,	C: Cytoplasm	CB: Cytoplasmic basophilia	CV: Cytoplasmic vacuoles)}
\label{tab:4}
\begin{tabular}{|c|ccccccc|}

\hline
\multirow{2}{*}{Method} & 
\multicolumn{1}{|c|}{NC} & \multicolumn{1}{|c|}{NS} & \multicolumn{1}{|c|}{N} & \multicolumn{1}{|c|}{C} & \multicolumn{1}{|c|}{CB} & \multicolumn{1}{c|}{CV} & \multicolumn{1}{|c|}{WBC} \\
\cline{2-8}
& \multicolumn{6}{|c|}{F1} & \multicolumn{1}{|c|}{mAP}   \\
\hline
CBM \cite{keser2023interpretable}& 21.9 & 96.2  & 41.8 & 77.2&  70.2 & 3.33  &\multicolumn{1}{|c|}{27.6}   \\
AttriDet & \textbf{73.9}  & \textbf{95.9}& \textbf{54.3}& \textbf{89.7}& \textbf{83.6}  &\textbf{29.1}  &\multicolumn{1}{|c|}{\textbf{28.2}}    \\
\hline
\end{tabular}
\end{table*}

\section{Conclusions}
In this paper, we have presented a large-scale WBC Leukemia dataset containing 12 subsets by using two different quality microscopes with multiple cameras at different resolutions (10x,40x,100x). The collected dataset contains 14 types of WBC-level localization with their seven distinct morphological attributes. Based on morphological information, we have provided an explainable AttriDet method to detect the WBC type with its morphological attributes. We believe that the presented dataset and proposed approach will facilitate future research in explainable, robust, and generalized Leukemia detection.

\section*{Acknowledgement}
We extend our sincere gratitude to Dr. Asma Saadia from Central Park Medical College, Lahore, and Dr. Ghulam Rasul from Ittefaq Hospital, Lahore, for their insightful discussions. Additionally, we express our appreciation to the dataset collection team, particularly Aurang Zaib, Nimra Dilawar, Sumayya Inayat, and Ehtasham Ul Haq, for their dedicated efforts. Furthermore, we acknowledge Google for their partial funding support for this project.


%
\newpage
\bibliographystyle{splncs04}
\bibliography{manuscriptV1}

\begin{thebibliography}{10}
\providecommand{\url}[1]{\texttt{#1}}
\providecommand{\urlprefix}{URL }
\providecommand{\doi}[1]{https://doi.org/#1}

\bibitem{aria2021acute}
Aria, M., Ghaderzadeh, M., Bashash, D., Abolghasemi, H., Asadi, F., Hosseini, A.: Acute lymphoblastic leukemia (all) image dataset. Kaggle  (2021)

\bibitem{azad2015short}
Azad, M., Biniaz, R.B., Goudarzi, M., Mobarra, N., Alizadeh, S., Nasiri, H., Fard, A.D., Kaviani, S., Moghadasi, M.H., Sarookhani, M.R., et~al.: Short view of leukemia diagnosis and treatment in iran. International journal of hematology-oncology and stem cell research  \textbf{9}(2), ~88 (2015)

\bibitem{bodzas2023high}
Bodzas, A., Kodytek, P., Zidek, J.: A high-resolution large-scale dataset of pathological and normal white blood cells. Scientific Data  \textbf{10}(1), ~466 (2023)

\bibitem{chhikara2023global}
Chhikara, B.S., Parang, K.: Global cancer statistics 2022: the trends projection analysis. Chemical Biology Letters  \textbf{10}(1),  451--451 (2023)

\bibitem{fischler1981random}
Fischler, M.A., Bolles, R.C.: Random sample consensus: a paradigm for model fitting with applications to image analysis and automated cartography. Communications of the ACM  \textbf{24}(6),  381--395 (1981)

\bibitem{gbenjo2023leukemia}
Gbenjo, J.T., McCrary, G.L., Wilson, S.E.: Leukemia: What primary care physicians need to know. American Family Physician  \textbf{107}(4),  397--405 (2023)

\bibitem{gupta2019all}
Gupta, A., Gupta, R.: All challenge dataset of isbi 2019 [data set]. The Cancer Imaging Archive  (2019)

\bibitem{keser2023interpretable}
Keser, M., Schwalbe, G., Nowzad, A., Knoll, A.: Interpretable model-agnostic plausibility verification for 2d object detectors using domain-invariant concept bottleneck models. In: Proceedings of the IEEE/CVF Conference on Computer Vision and Pattern Recognition. pp. 3890--3899 (2023)

\bibitem{kouzehkanan2022large}
Kouzehkanan, Z.M., Saghari, S., Tavakoli, S., Rostami, P., Abaszadeh, M., Mirzadeh, F., Satlsar, E.S., Gheidishahran, M., Gorgi, F., Mohammadi, S., et~al.: A large dataset of white blood cells containing cell locations and types, along with segmented nuclei and cytoplasm. Scientific reports  \textbf{12}(1), ~1123 (2022)

\bibitem{labati2011all}
Labati, R.D., Piuri, V., Scotti, F.: All-idb: The acute lymphoblastic leukemia image database for image processing. In: 2011 18th IEEE international conference on image processing. pp. 2045--2048. IEEE (2011)

\bibitem{lowe2004distinctive}
Lowe, D.G.: Distinctive image features from scale-invariant keypoints. International journal of computer vision  \textbf{60},  91--110 (2004)

\bibitem{mallya2022deep}
Mallya, M., Hamarneh, G.: Deep multimodal guidance for medical image classification. In: International Conference on Medical Image Computing and Computer-Assisted Intervention. pp. 298--308. Springer (2022)

\bibitem{manescu2022automated}
Manescu, P., Narayanan, P., Bendkowski, C., Elmi, M., Claveau, R., Pawar, V., Brown, B.J., Shaw, M., Rao, A., Fernandez-Reyes, D.: Automated detection of acute promyelocytic leukemia in blood films and bone marrow aspirates with annotation-free deep learning. arXiv preprint arXiv:2203.10626  (2022)

\bibitem{manescu2023detection}
Manescu, P., Narayanan, P., Bendkowski, C., Elmi, M., Claveau, R., Pawar, V., Brown, B.J., Shaw, M., Rao, A., Fernandez-Reyes, D.: Detection of acute promyelocytic leukemia in peripheral blood and bone marrow with annotation-free deep learning. Scientific Reports  \textbf{13}(1), ~2562 (2023)

\bibitem{matek2019single}
Matek, C., Schwarz, S., Marr, C., Spiekermann, K.: A single-cell morphological dataset of leukocytes from aml patients and non-malignant controls (aml-cytomorphology\_lmu). The Cancer Imaging Archive (TCIA)[Internet]  (2019)

\bibitem{mekhalfi2023daca}
Mekhalfi, M.L., Boscaini, D., Poiesi, F.: {Detect, Augment, Compose, and Adapt: Four Steps for Unsupervised Domain Adaptation in Object Detection}. In: BMVC (2023)

\bibitem{rezatofighi2011automatic}
Rezatofighi, S.H., Soltanian-Zadeh, H.: Automatic recognition of five types of white blood cells in peripheral blood. Computerized Medical Imaging and Graphics  \textbf{35}(4),  333--343 (2011)

\bibitem{ridnik2021asymmetric}
Ridnik, T., Ben-Baruch, E., Zamir, N., Noy, A., Friedman, I., Protter, M., Zelnik-Manor, L.: Asymmetric loss for multi-label classification. In: Proceedings of the IEEE/CVF International Conference on Computer Vision. pp. 82--91 (2021)

\bibitem{rugwizangoga2022experience}
Rugwizangoga, B., Niyikora, N., Musabyimana, A., Izimukwiye, A.I., Aurelius, J., Martner, A., Umubyeyi, A.: Experience and perception of patients and healthcare professionals on acute leukemia in rwanda: A qualitative study. Cancer Management and Research pp. 1923--1934 (2022)

\bibitem{scotti2005automatic}
Scotti, F.: Automatic morphological analysis for acute leukemia identification in peripheral blood microscope images. In: CIMSA. 2005 IEEE International Conference on Computational Intelligence for Measurement Systems and Applications, 2005. pp. 96--101. IEEE (2005)

\bibitem{shah2021automated}
Shah, A., Naqvi, S.S., Naveed, K., Salem, N., Khan, M.A., Alimgeer, K.S.: Automated diagnosis of leukemia: a comprehensive review. IEEE Access  \textbf{9},  132097--132124 (2021)

\bibitem{sultani2022towards}
Sultani, W., Nawaz, W., Javed, S., Danish, M.S., Saadia, A., Ali, M.: Towards low-cost and efficient malaria detection. In: 2022 IEEE/CVF Conference on Computer Vision and Pattern Recognition (CVPR). pp. 20655--20664. IEEE (2022)

\bibitem{sun2021sparse}
Sun, P., Zhang, R., Jiang, Y., Kong, T., Xu, C., Zhan, W., Tomizuka, M., Li, L., Yuan, Z., Wang, C., et~al.: Sparse r-cnn: End-to-end object detection with learnable proposals. In: Proceedings of the IEEE/CVF conference on computer vision and pattern recognition. pp. 14454--14463 (2021)

\bibitem{tian2019fcos}
Tian, Z., Shen, C., Chen, H., He, T.: Fcos: Fully convolutional one-stage object detection. In: Proceedings of the IEEE/CVF international conference on computer vision. pp. 9627--9636 (2019)

\bibitem{tsutsui2024wbcatt}
Tsutsui, S., Pang, W., Wen, B.: Wbcatt: A white blood cell dataset annotated with detailed morphological attributes. Advances in Neural Information Processing Systems  \textbf{36} (2024)

\bibitem{ultralytics2021yolov5}
Ultralytics: {YOLOv5}: {A} state-of-the-art real-time object detection system. \url{https://docs.ultralytics.com} (2021), accessed: 7 Jan, 2024

\bibitem{walkovich2022disorders}
Walkovich, K., Connelly, J.A.: Disorders of white blood cells. In: Lanzkowsky's Manual of Pediatric Hematology and Oncology, pp. 207--235. Elsevier (2022)

\bibitem{zhang2022dino}
Zhang, H., Li, F., Liu, S., Zhang, L., Su, H., Zhu, J., Ni, L., Shum, H.: Dino: Detr with improved denoising anchor boxes for end-to-end object detection. arxiv 2022. arXiv preprint arXiv:2203.03605  (2022)

\bibitem{zhang2022rcmnet}
Zhang, R., Han, X., Lei, Z., Jiang, C., Gul, I., Hu, Q., Zhai, S., Liu, H., Lian, L., Liu, Y., et~al.: Rcmnet: A deep learning model assists car-t therapy for leukemia. Computers in Biology and Medicine  \textbf{150},  106084 (2022)

\end{thebibliography}
%




\end{document}


\title{Supplementary Materials}
%
%
\author{
Paper ID:4180
}
\authorrunning{Author et al.}
\maketitle 
\small
\begin{table}[h]
\centering
\small
\renewcommand{\arraystretch}{1.1}
\caption{LeukemiaAttri dataset details with its multi-domain subsets}
\label{tab:dataset}
\begin{tabular}{|c|l|l|c|l|c|c|c|}

\hline
Sr. \# & Dataset & Microscope & Resolution & Camera & Image & WBC& Artifacts \\
 & Name  & &  &  & Count & Count & \\
\hline
1 & H\_100x\_C1\hspace{0.1em} & OlympusCX23  & 100x & Honor 9X lite & 2416 & 7857 & 2560 \\
2 & H\_100x\_C2 & OlympusCX23 \hspace{0.1em} & 100x & HD1500T & 2416 & 7857 & 2560 \\
3 & H\_40x\_C1 & OlympusCX23 & 40x & Honor 9X lite \hspace{0.1em} & 2416 & 7857 & 2560 \\
4 & H\_40x\_C2 & OlympusCX23 & 40x & HD1500T & 2416 & 7857 & 2560 \\
5 & H\_10x\_C1 & OlympusCX23 & 10x & Honor 9X lite & 2416 & 7857 & 2560 \\
6 & H\_10x\_C2 & OlympusCX23 & 10x & HD1500T & 2416 & 7857 & 2560 \\
7 & L\_100x\_C1 & XSZ\_107BN & 100x & Honor 9X lite & 2416 & 4862 & 1120 \\
8 & L\_100x\_C2 & XSZ\_107BN & 100x & ZZCAT-5MP & 2416 & 4862 & 1120 \\
9 & L\_40x\_C1 & XSZ\_107BN & 40x & Honor 9X lite & 2416 & 7857 & 2560 \\
10 & L\_40x\_C2 & XSZ\_107BN & 40x & ZZCAT-5MP & 2416 & 7857 & 2560 \\
11 & L\_10x\_C1 & XSZ\_107BN & 10x & Honor 9X lite & 2416 & 7857 & 2560 \\
12 & L\_10x\_C2 & XSZ\_107BN & 10x & ZZCAT-5MP & 2416 & 7857 & 2560 \\
\hline
\multicolumn{5}{|c|}{Total Count}  & 28,992 & 88,294 & 27,840 \\
\hline
\end{tabular}
\end{table}

\begin{figure}[!h]
    \includegraphics[width=\linewidth]{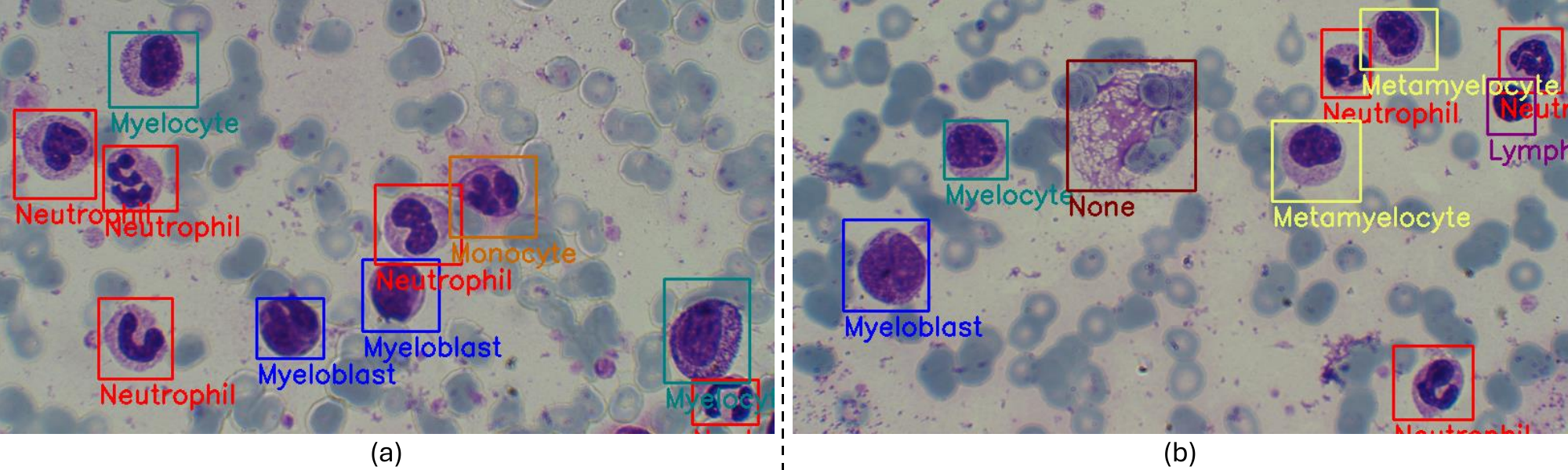}

    \caption{Two samples images from the LeukemiaAttri dataset are shown in (a) and (b) sections with multiple WBC types}
    \label{fig:1}
\end{figure}
%




\begin{figure}[!htbp]
 
   \begin{subfigure}[b]{0.45\textwidth}
   
    \includegraphics[width=\linewidth]{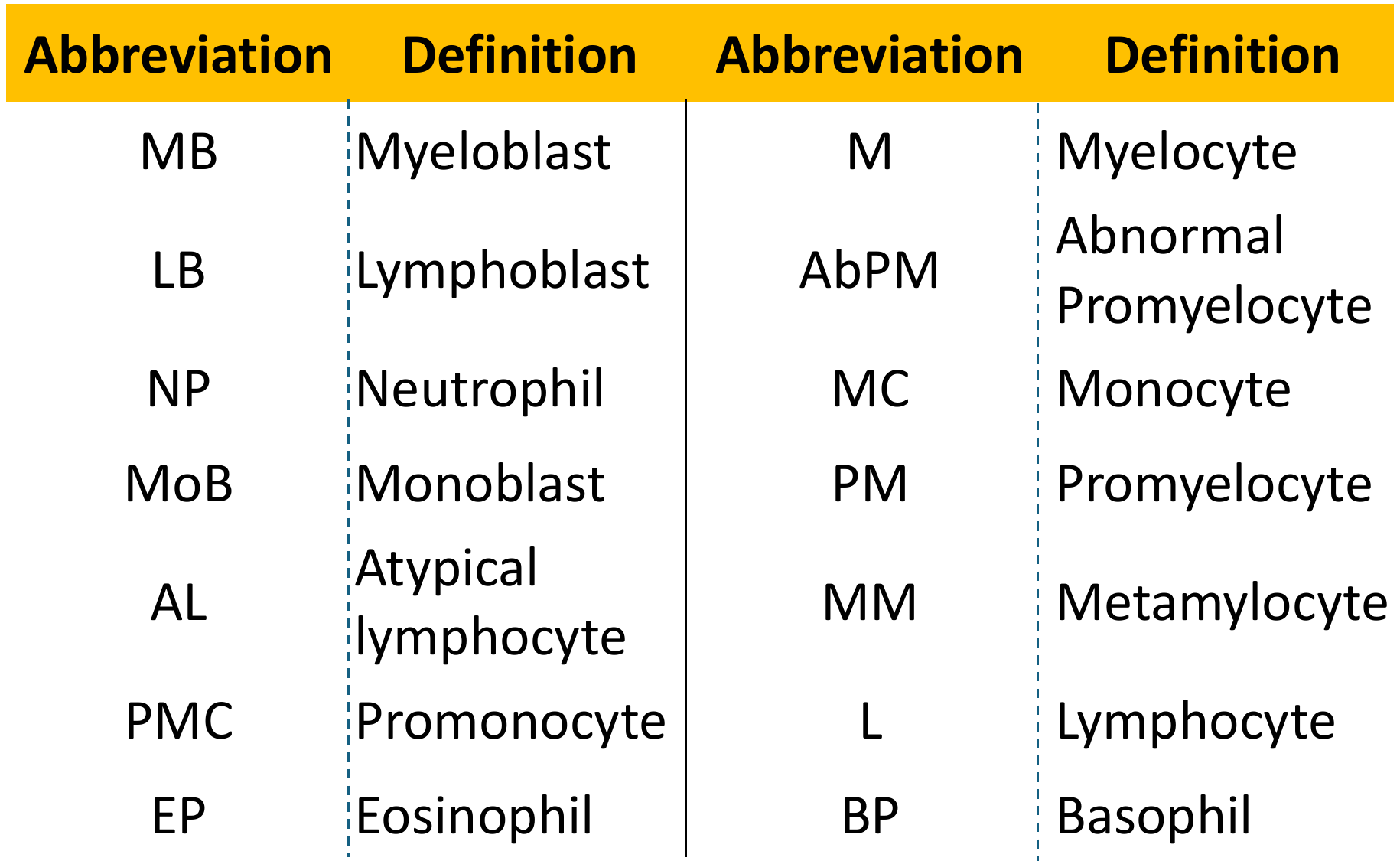}
    \caption{}
    \label{fig:Abbrivation}
  \end{subfigure}
\hfill
  \begin{subfigure}[b]{0.45\textwidth}
    \includegraphics[width=\linewidth]{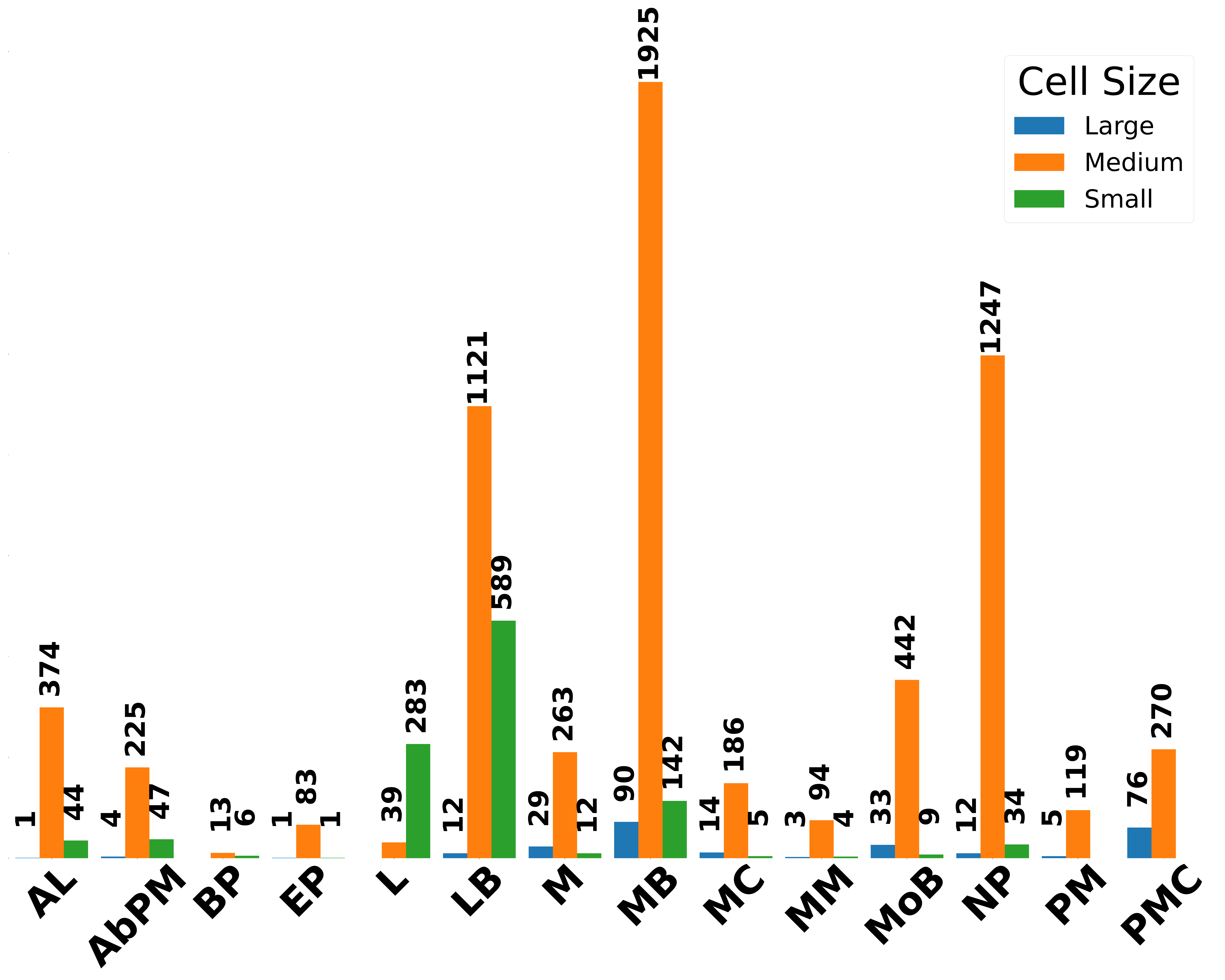}
    \caption{}
    \label{fig:Cell_Size}
  \end{subfigure}

  \vspace{\baselineskip} 

  \begin{subfigure}[b]{0.45\textwidth}
    \includegraphics[width=\linewidth]{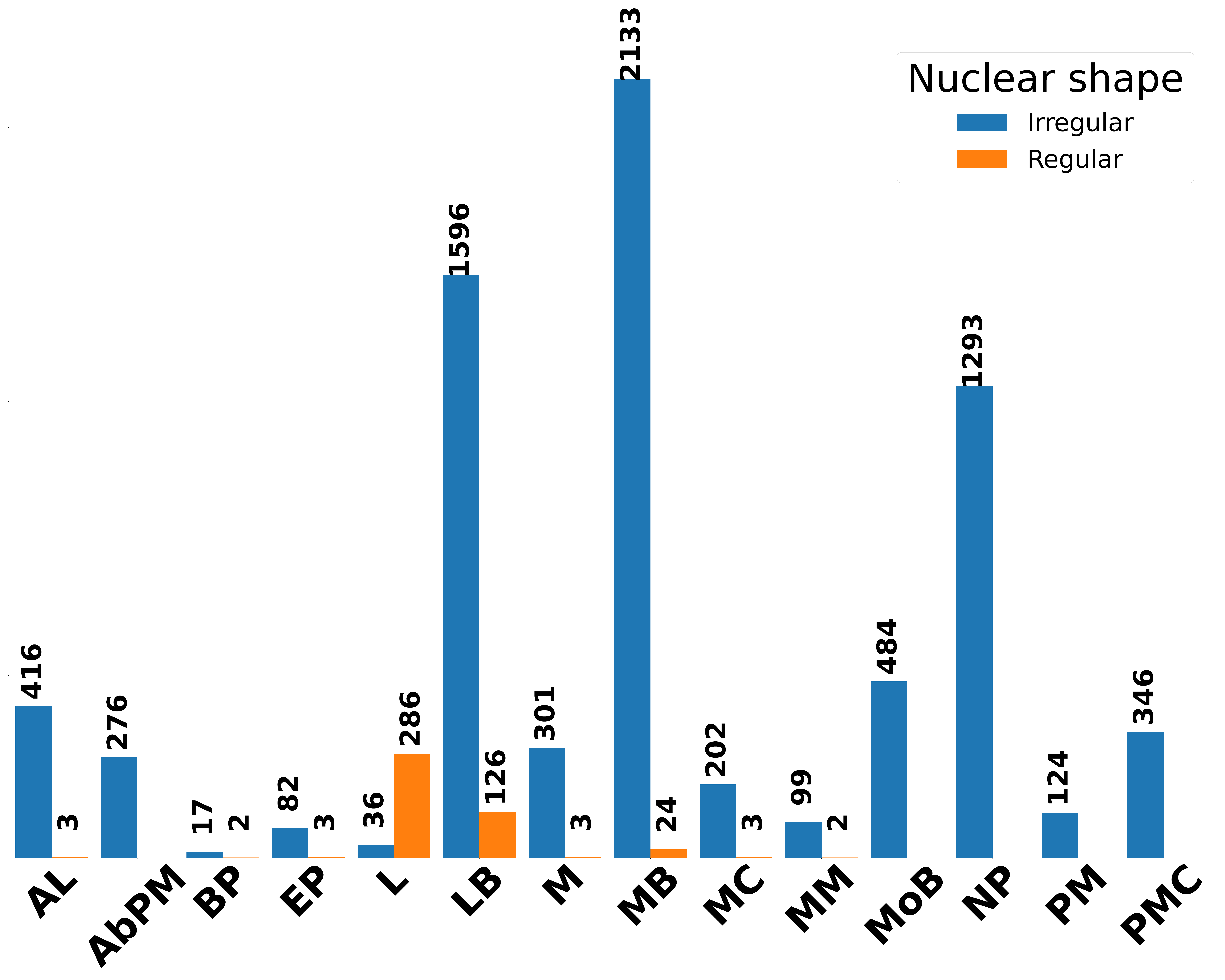}
    \caption{}
    \label{fig:Nuclear_shape}
  \end{subfigure}
  \hfill
  \begin{subfigure}[b]{0.45\textwidth}
    \includegraphics[width=\linewidth]{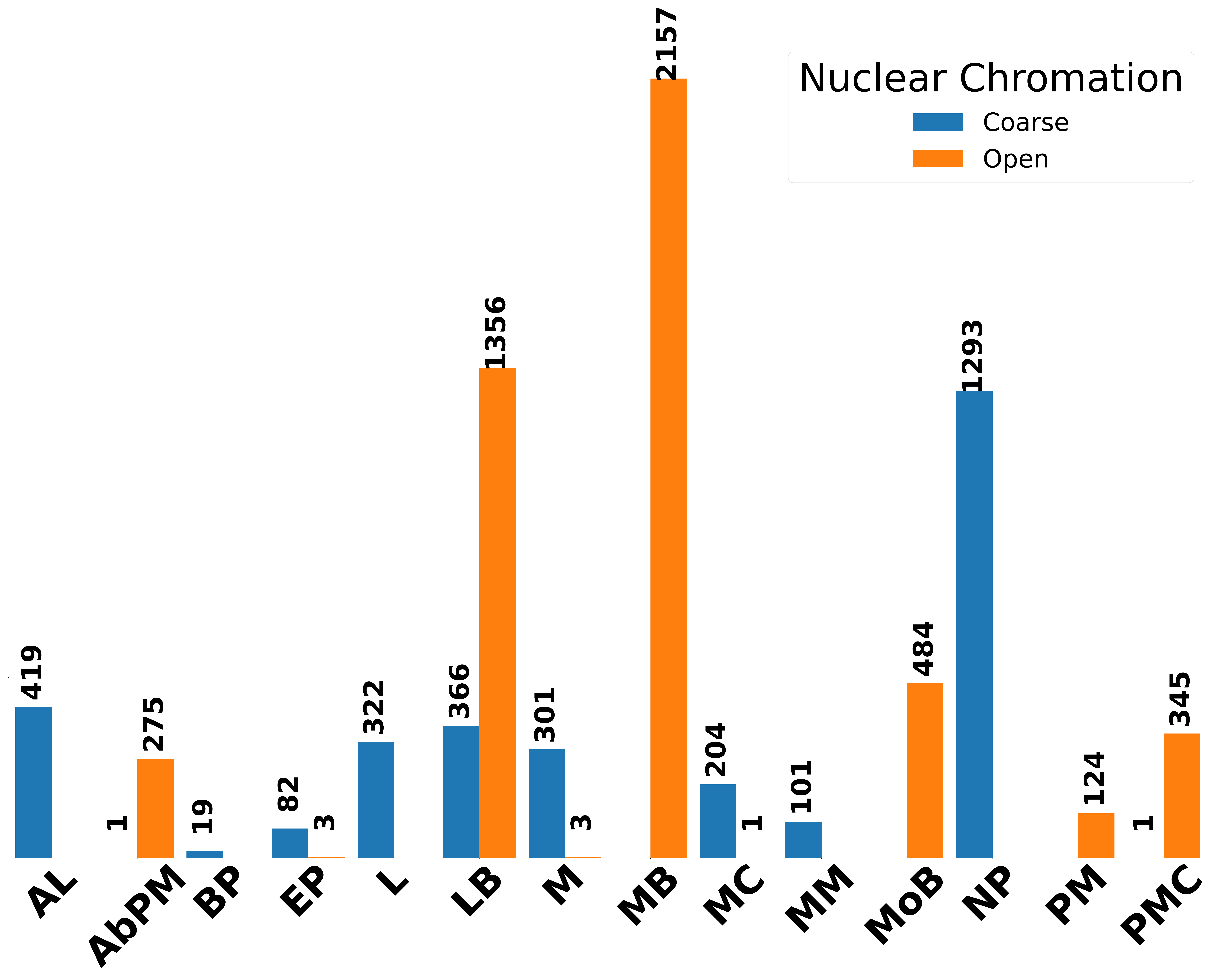}
    \caption{}
    \label{fig:Nuclear_Chromation}
  \end{subfigure}

  \vspace{\baselineskip} 

  \begin{subfigure}[b]{0.45\textwidth}
    \includegraphics[width=\linewidth]{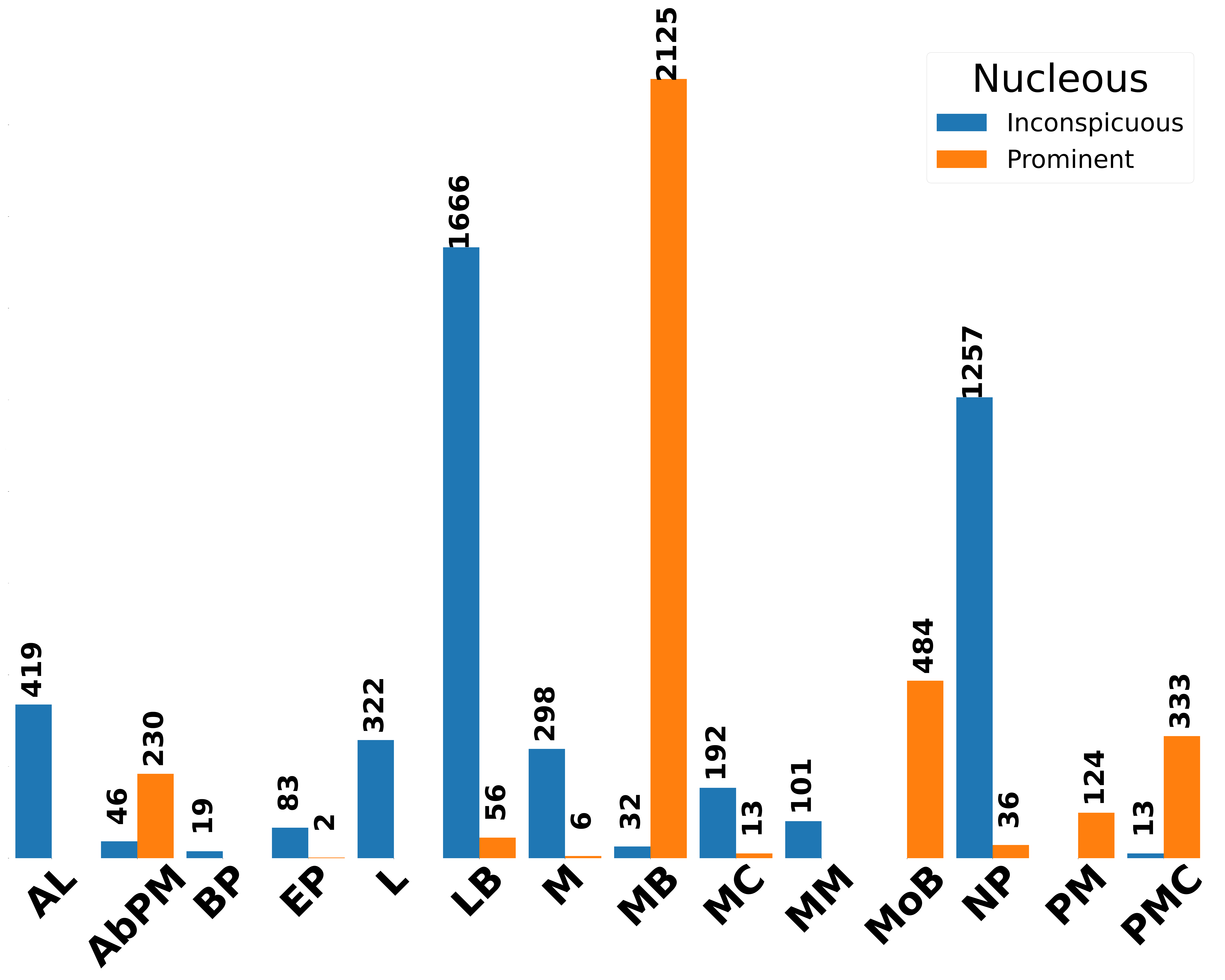}
    \caption{}
    \label{fig:Nucleous}
  \end{subfigure}
  \hfill
  \begin{subfigure}[b]{0.45\textwidth}
    \includegraphics[width=\linewidth]{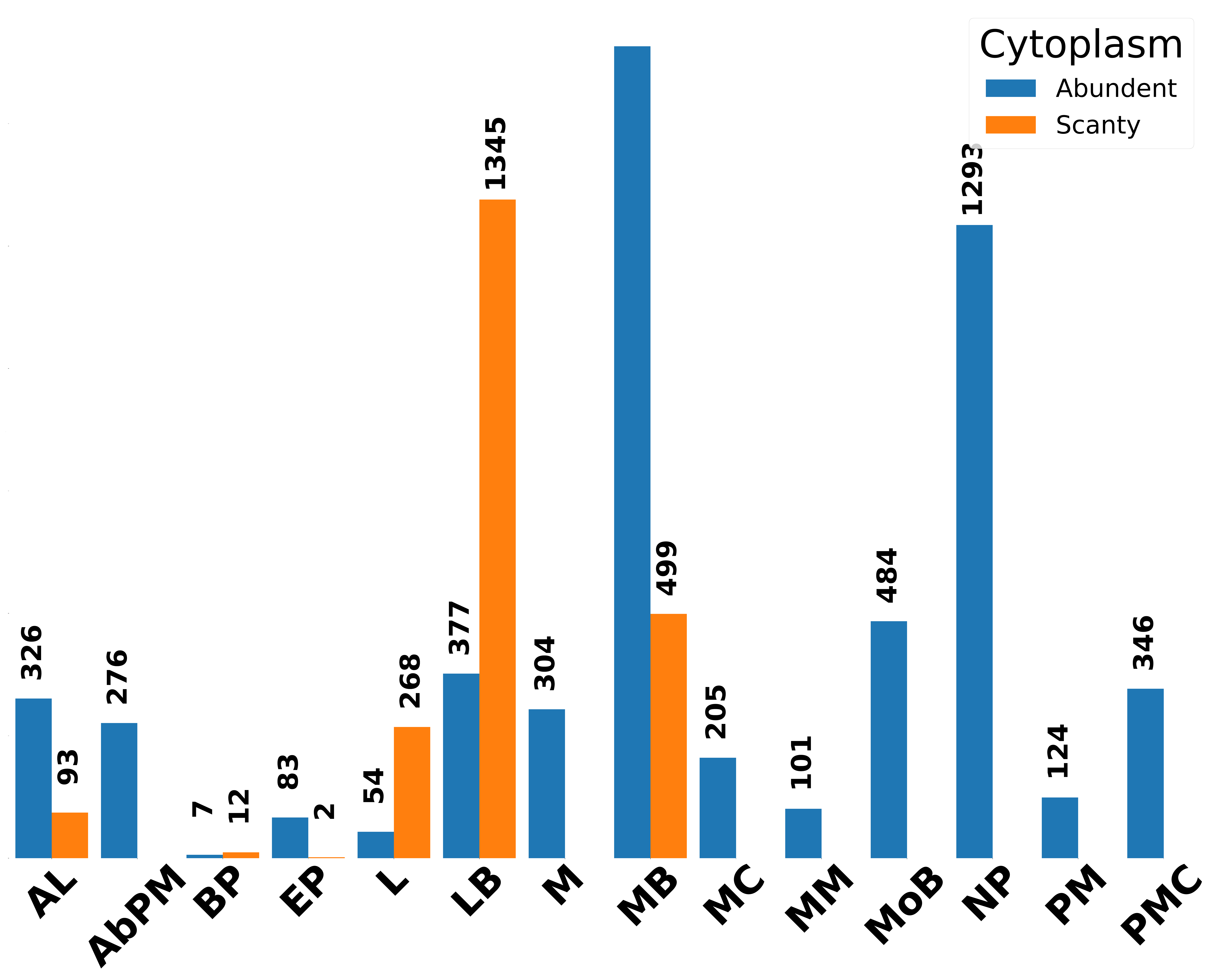}
    \caption{}
    \label{fig:Cytoplasm}
  \end{subfigure}

  \vspace{\baselineskip} 

  \begin{subfigure}[b]{0.45\textwidth}
    \includegraphics[width=\linewidth]{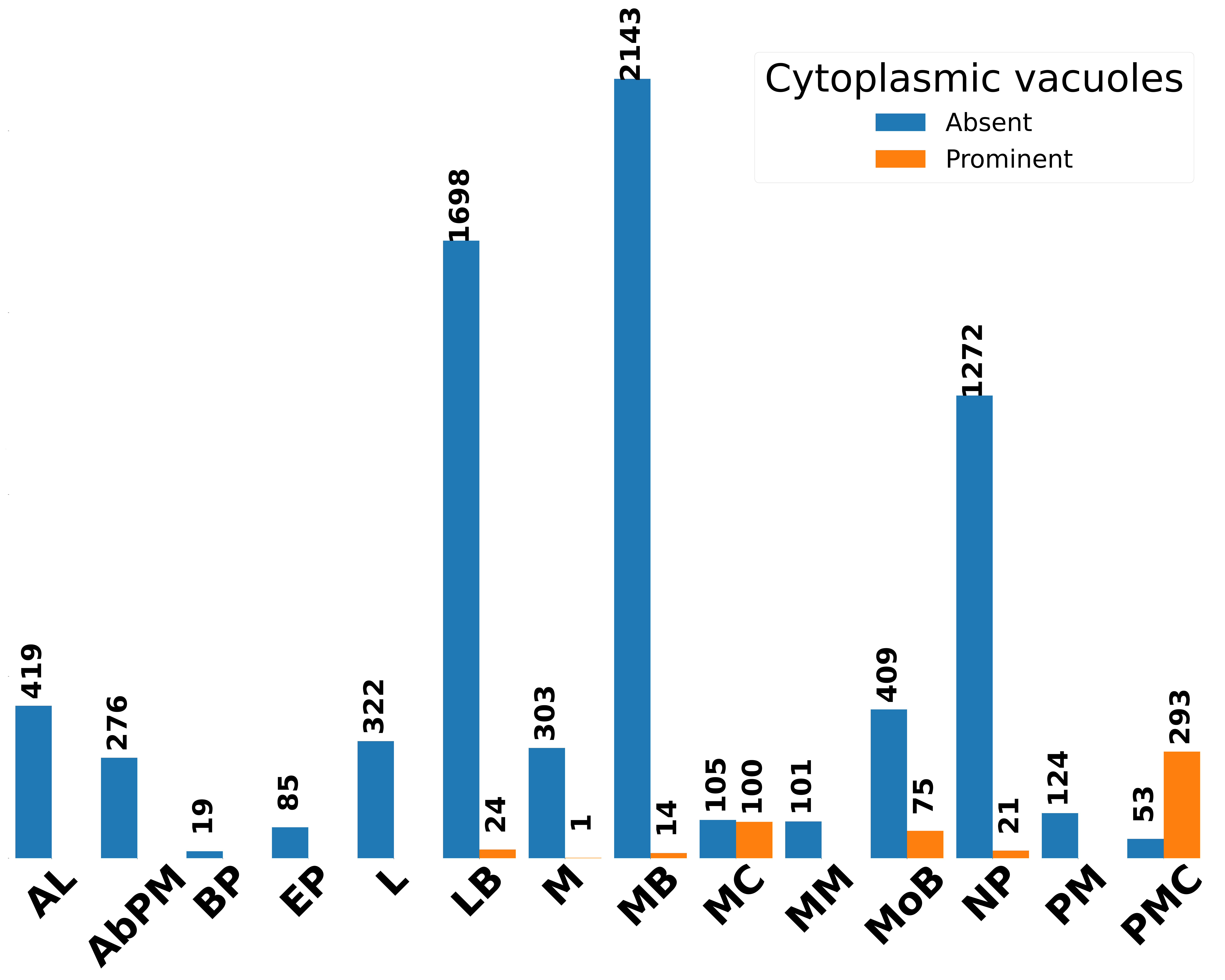}
    \caption{}
    \label{fig:Cytoplasmic_vacuoles}
  \end{subfigure}
  \hfill
  \begin{subfigure}[b]{0.45\textwidth}
    \includegraphics[width=\linewidth]{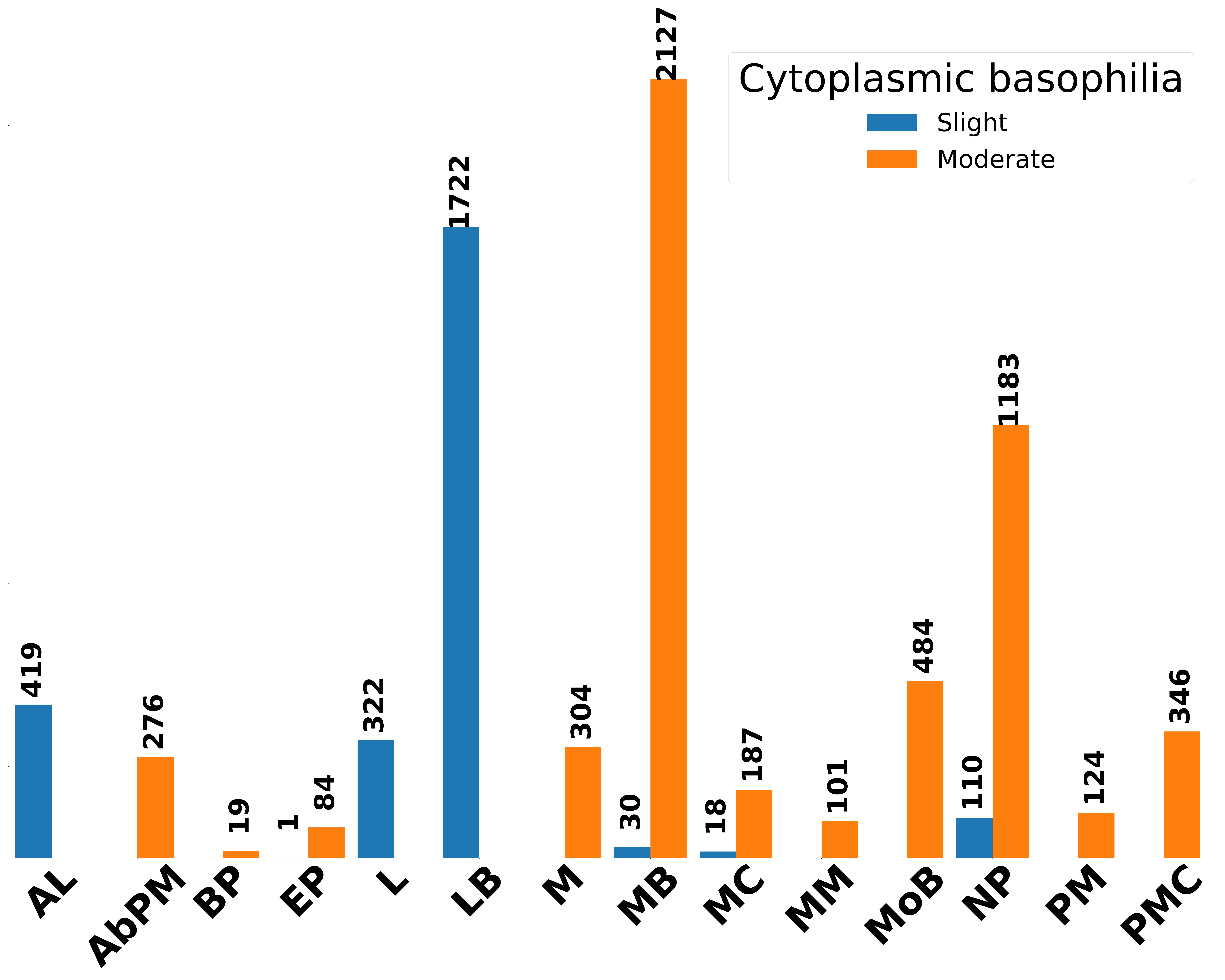}
    \caption{}
    \label{fig:Cytoplasmic_basophilia}
  \end{subfigure}

  \vspace{\baselineskip} 

  \caption{Attribute label counts w.r.t WBC types in LeukemiaAttri dataset: (a) Abbreivations of WBC types, (b) Cell Size, (c) Nuclear Shape, 
  (d) Nuclear Chromatin, (e) Nucleus, (f) Cytoplasm, (g) Cytoplasmic vacuoles, (h) Cytoplasmic basophilia
  }
  \label{fig:Combined_Figure}
\end{figure}